\documentclass[11pt,a4paper]{article}
\usepackage[hyperref]{naaclhlt2019}
\usepackage{times}
\usepackage{latexsym}

\usepackage{url}

\usepackage[T1]{fontenc}    
\usepackage{booktabs}       
\usepackage{amsfonts}       
\usepackage{nicefrac}       
\usepackage{microtype}      
\usepackage{helvet}  
\usepackage{courier}  
\usepackage{url}  
\usepackage{graphicx}  

\usepackage{amsfonts}       
\usepackage{amsmath}
\usepackage{amssymb}
\usepackage{bbm}
\usepackage{amsthm}
\usepackage{bm}
\usepackage{booktabs}       
\usepackage[T1]{fontenc}    
\usepackage{hyperref}       
\usepackage[utf8]{inputenc} 
\usepackage{microtype}      
\usepackage{multirow}
\usepackage{nicefrac}       
\usepackage{subcaption}

\aclfinalcopy 
\def\aclpaperid{133} 

\title{Personalized Neural Embeddings for Collaborative Filtering with Text}

\author{Guangneng Hu \\
  Department of Computer Science and Engineering \\
  Hong Kong University of Science and Technology, Hong Kong, China \\
  {\tt njuhgn@gmail.com} \\
}

\date{}

\begin{document}

\maketitle

\begin{abstract}
Collaborative filtering (CF) is a core technique for recommender systems. Traditional CF approaches exploit user-item relations (e.g., clicks, likes, and views) only and hence they suffer from the data sparsity issue. Items are usually associated with unstructured text such as article abstracts and product reviews. We develop a Personalized Neural Embedding (PNE) framework to exploit both interactions and words seamlessly. We learn such embeddings of users, items, and words jointly, and predict user preferences on items based on these learned representations. PNE estimates the probability that a user will like an item by two terms---behavior factors and semantic factors. On two real-world datasets, PNE shows better performance than four state-of-the-art baselines in terms of three metrics. We also show that PNE learns meaningful word embeddings by visualization.
\end{abstract}

\section{Introduction}

Recommender systems are widely used in e-commerce platforms,
such as to help consumers buy products at Amazon, watch videos on Youtube, and read articles on Google News. They are useful to alleviate the information overload and improve user satisfaction. Given history records of consumers such as the product transactions and movie watching, collaborative filtering (CF) is among the most effective approaches based on the simple intuition that if users rated items similarly in the past then they are likely to rate items similarly in the future~\cite{sarwar2001item}.

History records include both implicit (e.g., purchase and clicks) and explicit (e.g., likes/dislikes and ratings) feedback which can be represented as a user-item interaction matrix. Typically, observed user-item interactions are incomplete with a large portion remaining not recorded. The goal of recommendation is to predict user preferences on these missing interactions. This setting requires to complete the partial observed rating matrix. Matrix Factorization (MF) techniques which can learn latent factors for users and items are the main cornerstone for CF~\cite{mnih2008pmf,koren2008mf,koren2009matrix}. It is effective and flexible to integrate with additional data sources~\cite{MR3}. Recently, neural networks like Multilayer Perceptron (MLP) are used to learn an interaction function from data with the power of learning highly nonlinear relationships between users and items~\cite{dziugaite2015neural,cheng2016wide,he2017neural,CoNet}. MF and neural CF exploit user-item behavior interactions only and hence they both suffer from the data sparsity and cold-start issues.

Items are usually associated with unstructured text, like news articles and product reviews. These additional sources are essential for recommendation beyond user-item interactions since they contain independent and diverse information. Hence, they provide an opportunity to alleviate the data sparsity issue~\cite{ganu2009review,MR3PP}. For application domains like recommending research papers and news articles, the unstructured text associated with the item is its text content~\cite{wang2011ctr,wang2015cdl,bansal2016gru}. For some domains like recommending products, the unstructured text associated with the item is its user reviews which justify the rating behavior~\cite{mcauley2011hft,he2016vbpr}. These methods adopt topic modelling techniques and neural networks to exploit the item content leading to performance improvement.

A typical way of exploiting text content is to firstly extract a feature vector for each document by averaging word embeddings in the document, and then to learn a text factor corresponding to this feature vector~\cite{hu2017tbpr}. These embeddings are pre-trained from a large corpus such as Wikipedia. This approach separates the extraction of text feature from the learning of user-item interaction. These two processes cannot benefit from each other and errors in the previous step maybe propagate to the successive steps. Another way is to learn a topic vector using topic modelling~\cite{wang2011ctr,mcauley2011hft,bao2014topicmf} by aligning behavior factors and topic factors with a link function such as softmax and offset.

Recently, neural networks are used to learn a representation from the text using autoencoders~\cite{wang2015cdl,zhang2016cke}, recurrent networks~\cite{bansal2016gru}, and convolutional networks~\cite{zheng2017joint,catherine2017transnets}. These methods treat different words in the document as equal importance and do not match word semantics with the specific user. Instead, we achieve to learn a personalized word embedding with the guidance of user-item interactions. That is, the importance of words is learned to match user preferences. The attention mechanism can be used to learn these importance weights. Memory Networks (MemNet) have been used in recommendation to model item content~\cite{hu2018lcmr,huang2017mention}, capture user neighborhood~\cite{ebesu2018memory}, and learn latent relationships~\cite{tay2018latent}. We follow this thread to adapt a MemNet to match word semantics with user preferences.

In this paper, we propose a novel neural framework to exploit relational interactions and text content seamlessly. The proposed Personalized Neural Embedding (PNE) model fuses semantic representations learnt from unstructured text with behavior representations learnt from user-item interactions jointly for effective estimation of user preferences on items. PNE estimates the preference probability by two kinds of factors. The {\it behavior factor} is to capture the personalized preference of a user to an item learned from behavior interactions. The {\it semantic factor} is to capture the high-level representation attentively extracted from the unstructured text by matching word semantics with user preferences.

To model the behavior factor, we adopt a neural CF approach, which learns the user-item nonlinear interaction relationships using a neural network (CFNet). To model the semantic factor, we adopt a memory network to match word semantics with the specific user via the attention mechanism inherent in the memory module (MemNet), determining which words are highly relevant to the user preferences. PNE integrates relational interactions with unstructured text by bridging neural CF and memory networks. PNE can also learn meaningful word embeddings.

\section{Approach}

We present PNE to jointly learn representations of users, items, and words. PNE seamlessly captures nonlinear user-item interaction relationships and matches word semantics with user preferences.

Denote the set of users by $\mathcal{U}$ and items by $\mathcal{I}$. We use a rating matrix $\bm{Y} \in \mathbb{R}^{|\mathcal{U}| \times |\mathcal{I}|}$ to describe user-item interactions where each entry $y_{ui} \in \{0,1\}$ is 1 (observed entries) if user $u$ has an interaction with item $i$ and 0 (unobserved entries) otherwise. Usually the interaction matrix is very sparse since a user $u \in \mathcal{U}$ only consumed a very small subset of all items. For the task of item recommendation, each user is only interested in identifying $topK$ items (typically $K$ is small e.g. tens or hundreds). The items are ranked by their predicted scores:
\begin{equation}
\hat y_{ui} = f(u,i | \Theta),
\end{equation}
where $f$ is the interaction function and $\Theta$ denotes model parameters.

\subsection{Architecture}

\begin{table*}[]					
\centering
\caption{Datasets and statistics. }
\label{tb:data}			
\begin{tabular}{c | cccc | cc}
Dataset & \#user & \#item & \#rating & \#word    & \#density & avg. words \\
\hline
Amazon  & 8,514  & 28,262 & 56,050     & 1,845,387 & 0.023\%   & 65.3       \\
\hline
Cheetah  & 15,890 & 84,802 & 477,685    & 612,839   & 0.035\%   & 7.2 \\
\end{tabular}
\end{table*}

PNE consists of a CF network (CFNet) to learn a nonlinear interaction function and of a memory network (MemNet) to match word semantics with user preferences. The information flow in PNE goes from the input $(u,i)$ to the output $\hat{y}_{ui}$ through the following five modules.

1. {\it Input: $(u,i) \rightarrow \vec{e}_u,\vec{e}_i$} This module encodes user-item interaction indices. We adopt the one-hot encoding. It takes user $u$ and item $i$, and maps them into one-hot encodings $\vec{e}_u \in \{0,1\}^{|\mathcal{U}|}$ and $\vec{e}_i \in \{0,1\}^{|\mathcal{I}|}$ where only the element corresponding to that user/item index is 1 and all others are 0.

2. {\it Embedding: $\vec{e}_u,\vec{e}_i \rightarrow \bm{x}_{ui}$} This module firstly embeds one-hot encodings into continuous representations $\bm{x}_u=\bm{P}^T \vec{e}_u$ and $\bm{x}_i=\bm{Q}^T \vec{e}_i$ by embedding matrices $\bm{P} \in \mathbb{R}^{|\mathcal{U}| \times d}$ and $\bm{Q} \in \mathbb{R}^{|\mathcal{I}| \times d}$ respectively, where $d$ is the latent dimension. It then concatenates them as $\bm{x}_{ui} = [\bm{x}_u,\bm{x}_i]$ to be the input of following CFNet and MemNet modules.

3. {\it CFNet: $\bm{x}_{ui} \rightarrow \bm{z}_{ui}$} This module is a CF approach to exploit user-item interactions. It takes continuous representations from the embedding module and then transforms to a final {\it behavior factor} representation:
\begin{equation}
\bm{z}_{ui}= \textrm{ReLU}(\bm{W}\bm{x}_{ui} + \bm{b}),
\end{equation}
where $\textrm{ReLU}(x)=\max(0,x)$ is an activation function, and $\bm{W}$ and $\bm{b}$ are connection weights and biases.

4. {\it MemNet: $\bm{x}_{ui} \rightarrow \bm{o}_{ui}$} This module is to model the item content with the guidance of user-item interaction. The item content is modelled by memories. It takes  representations from both the embedding module and the review text $d_{ui}$ associated with the corresponding user-item $(u,i)$ into a final {\it semantic factor} representation:
\begin{equation}
\bm{o}_{ui}= \sum\nolimits_{j:\,w_j \in d_{ui}} \textrm{Softmax}(a_j^{u,i}) \bm{c}_j,
\end{equation}
where the external memory slot $\bm{c}_j$ is an embedding vector for word $w_j$ by mapping it with an external memory matrix $\bm{C}$. The attentive weight $a_j^{u,i}$ encodes the relevance of user $u$ to word $w_j$ by content-based addressing:
\begin{equation}\label{eq:attention}
a_j^{u,i} = \bm{x}_{ui}^T \bm{m}_j^{u,i},
\end{equation}
where memory $\bm{m}_j^{u,i}$ is concatenated from internal memory slots $\{\bm{m}_j^{u},\bm{m}_j^{i}\}$ which are mapped from word $w_j$ by internal memory matrices $\bm{A}^u$ for user attention and $\bm{A}^i$ for item attention.

5. {\it Output: $[\bm{z}_{ui}, \bm{o}_{ui}] \rightarrow \hat y_{ui}$} This module predicts the recommendation score $\hat y_{ui}$ for a given user-item pair based on the representation of both behavior factor and semantic factor from CFNet and MemNet respectively:
\begin{equation}
\tilde {\bm{z}}_{ui} = [\bm{z}_{ui}, \bm{o}_{ui}].
\end{equation}
The output is the probability that the input pair is a positive interaction. This is achieved by a logistic layer:
\begin{equation}
\hat y_{ui} =  \frac {1} {1 + \exp(-\bm{h}^T \tilde {\bm{z}}_{ui})},
\end{equation}
 where $\bm{h}$ is model parameter.

\subsection{Learning}

We adopt the binary cross-entropy loss:
\begin{equation}
\mathcal{L}(\Theta) =  - \sum\limits_{(u,i) \in \mathcal{S} } y_{ui} \log{\hat y_{ui}} + (1-y_{ui}) \log(1 - \hat y_{ui} ),
\end{equation}
where $\mathcal{S} = \bm{Y}^+ \cup \bm{Y}^-$ is the union of observed interactions and randomly sampled negative examples. Model parameters $\Theta = \{\bm{P},\bm{Q},\bm{W},\bm{b},\bm{A},\bm{h}\}$ where we use a single word embedding matrix $\bm{A}$ by sharing all memory matrices $\bm{A}^u,\bm{A}^i$, and $\bm{C}$ in order to reduce model complexity. The objective function can be optimized by stochastic gradient descent.

\begin{table}[]
\caption{Categorization of recommender approaches.}\label{tb:baseline}
\resizebox{0.48\textwidth}{!}{
\begin{tabular}{|c|c|c|}
\hline
 Baselines & Shallow  method & Deep method \\
\hline
 CF    & BPRMF        & MLP   \\
\hline
CF w/ Text     & HFT, TBPR  & LCMR, PNE (ours)\\
\hline
\end{tabular}
}
\end{table}

\section{Experiment}

In this section, we evaluate PNE on two datasets with five baselines in terms of three metrics.

\subsection{ Datasets }

We evaluate on two real-world datasets. The public {\bf Amazon} products ~\cite{mcauley2011hft} and a company {\bf Cheetah} Mobile news~\cite{hu2018lcmr,liu2018transferable} (see Table~\ref{tb:data}). We preprocess the data following the strategy in ~\cite{wang2011ctr}. The size of word vocabulary is 8,000.

\subsection{ Metrics and Baselines }

We adopt leave-one-out evaluation~\cite{CoNet} and use three ranking metrics: hit ratio (HR), normalized discounted cumulative gain (NDCG), and mean reciprocal rank (MRR).

We compare with five baselines (see Table~\ref{tb:baseline}).
\begin{itemize}
\item {BPR}~\cite{rendle2009bpr} is a latent factor model based on matrix factorization.
\item {HFT}~\cite{mcauley2011hft} adopts topic distributions to learn latent factors from text reviews.
\item {TBPR}~\cite{hu2017tbpr} extends BPR by integrating text content via word embedding features. The word embeddings used in the TBPR are pre-trained by GloVe~\cite{Pennington2014GloveGV}.
\item {MLP}~\cite{he2017neural} is a neural CF approach. Note that, CFNet in PNE is an MLP with only one hidden layer.
\item {LCMR}~\cite{hu2018lcmr} is a deep model for CF with unstructured text.
\end{itemize}

Our method is implemented by TensorFlow~\cite{abadi2016tensorflow}. Parameters are randomly initialized from Gaussian with optimizer Adam~\cite{Kingma2015AdamAM}. Learning rate is 0.001,  batch size is 128, the ratio of negative sampling is 1.

\begin{table}[]
\centering
\caption{Results ($\times 100$) on Amazon dataset. Best baseline marked with asterisk and best result in boldfaced. }
\label{table:results-amazon}
\resizebox{0.5\textwidth}{!}{	
\begin{tabular}{c c| rrr lll}
\hline \hline
\multirow{2}{*}{TopK}  & \multirow{2}{*}{Metric} & \multicolumn{6}{c}{Method}  \\
\cline{3-8}
                         &      & BPR  & HFT    &TBPR &  MLP   & LCMR   & PNE       \\
\hline \hline
\multirow{3}{*}{5}  & HR   & 8.10 & 10.77 & 15.17 & 21.00*& 20.24 & {\bf 23.52} \\
                         & NDCG & 5.83 & 8.15 & 12.08 & 14.86*& 14.51 & {\bf 16.46}  \\
                         & MRR  & 5.09 & 7.29 & 11.04 & 12.83*& 12.63 & {\bf 14.13}  \\
\hline
\multirow{3}{*}{10} & HR   & 12.04 & 13.60 & 17.77 & 28.36*& 28.36*& {\bf 31.86}  \\
                         & NDCG & 7.10 & 9.07 & 12.91 & 16.97*& 16.78 & {\bf 19.15}  \\
                         & MRR  & 5.61 & 7.67 & 11.38 & 13.71*& 13.56 & {\bf 15.24}  \\
\hline
\multirow{3}{*}{20} & HR   & 18.21 & 27.82 & 22.68 & 38.20 & 39.51*& {\bf 42.21}  \\
                         & NDCG & 8.64 & 12.52 & 14.14 & 18.99 & 19.18*& {\bf 21.75} \\
                         & MRR  & 6.02 & 8.54 & 11.71 & 14.26*& 14.20 & {\bf 15.95}  \\
\hline \hline
\end{tabular}
}
\end{table}

\begin{table}[]
\centering
\caption{Results ($\times 100$) on Cheetah dataset. Best baseline marked with asterisk and best result in boldfaced. }
\label{table:results-mobile}
\resizebox{0.5\textwidth}{!}{	
\begin{tabular}{c c| rr llll }
\hline \hline
\multirow{2}{*}{TopK}  & \multirow{2}{*}{Metric} & \multicolumn{6}{c}{Method}  \\
\cline{3-8}
                         &      & BPR  & HFT    &TBPR &  MLP   & LCMR   & PNE        \\
\hline \hline
\multirow{3}{*}{5}  & HR   & 43.80 & 49.66 & 49.48 & 53.80 & 54.76*& {\bf 56.48}   \\
                         & NDCG & 39.71 & 36.17 & 42.98*& 41.21 & 41.89 & {\bf 43.45}   \\
                         & MRR  & 36.06 & 31.75 & 38.26*& 37.02 & 37.62 & {\bf 39.11}   \\
\hline
\multirow{3}{*}{10} & HR   & 49.41 & 55.80 & 54.66 & 61.76 & 63.11*& {\bf 64.24} \\
                         & NDCG & 41.82 & 40.93 & 44.99*& 43.81 & 44.60 & {\bf 45.98}   \\
                         & MRR  & 36.94 & 33.65 & 39.13*& 38.10 & 38.74 & {\bf 40.16}  \\
\hline
\multirow{3}{*}{20} & HR   & 53.98 & 65.47 & 61.23 & 67.93 & 69.27*& {\bf 69.52}  \\
                         & NDCG & 43.16 & 43.79 & 46.82*& 45.29 & 46.19 & {\bf 47.32}  \\
                         & MRR  & 37.30 & 34.45 & 39.58*& 38.51 & 39.18 & {\bf 40.53}   \\
\hline \hline
\end{tabular}
}
\end{table}

\begin{figure}[]
\centering
\includegraphics[width=6.2cm, height=5.cm]{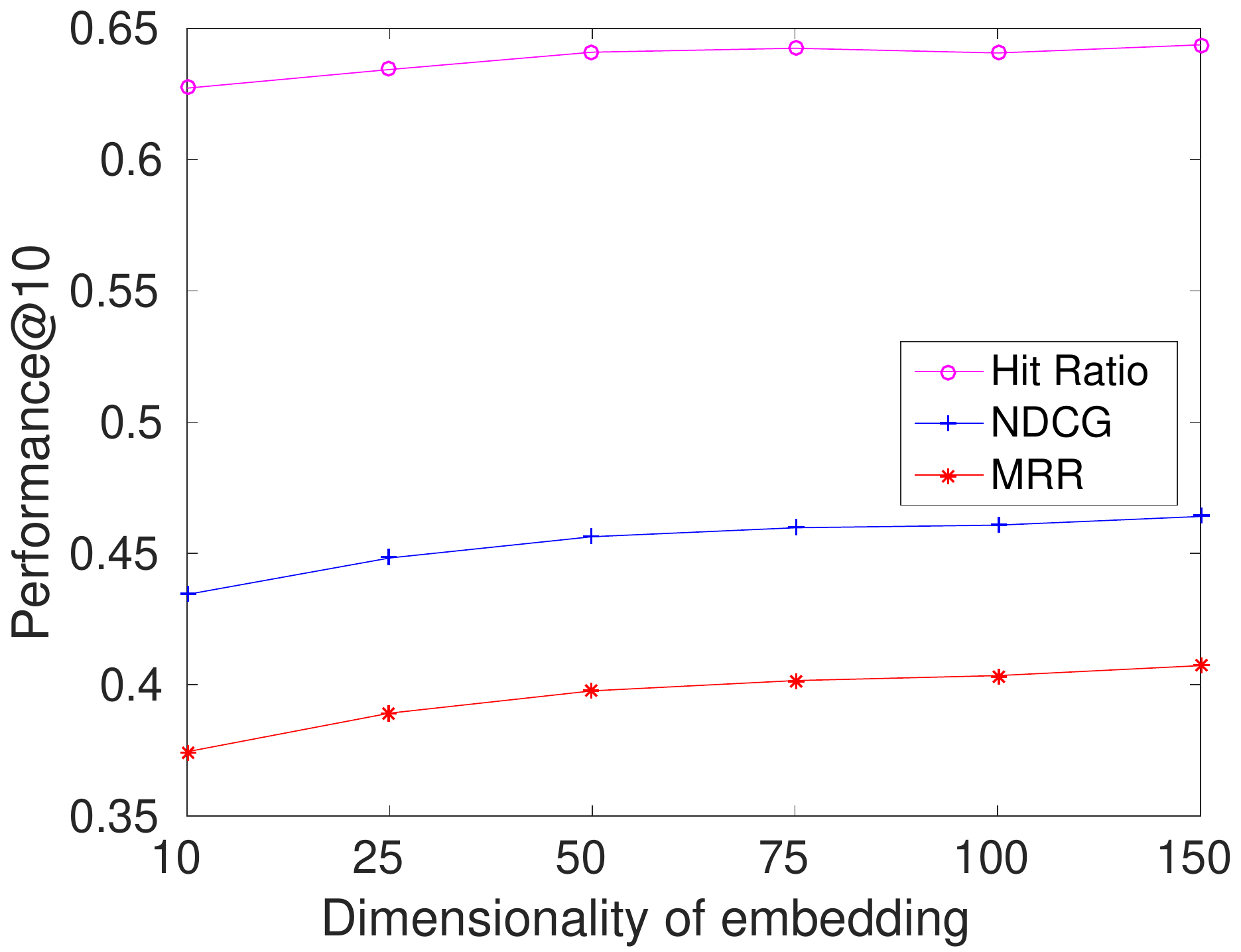}
\caption{Dimension of embedding. }
\label{fig:embedding}
\end{figure}

\begin{figure}[]
\centering
\includegraphics[width=6.2cm, height=5.cm]{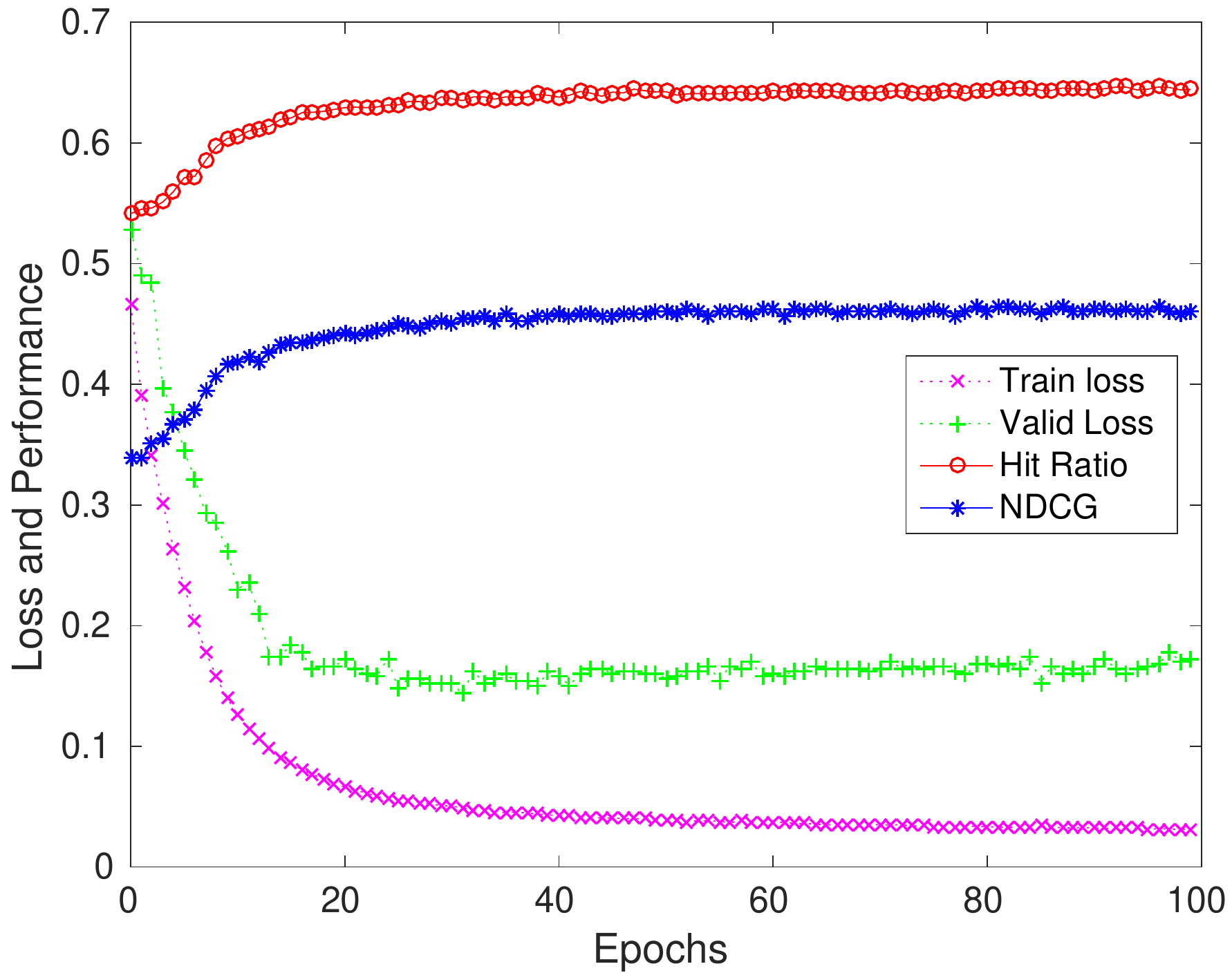}
\caption{Loss and performance@10.}
\label{fig:performance-loss-epoch}
\end{figure}

\subsection{ Results }

Results on two datasets are shown in Table~\ref{table:results-amazon} and Table~\ref{table:results-mobile}. We have some observations. First, PNE outperforms the neural CF method MLP on two datasets in terms of three ranking metrics. On Amazon dataset, PNE obtains a large improvement in performance gain with relative 12.3\% HR@10, 7.7\% NDCG@10, and 6.2\% MRR@5. On Cheetah Mobile dataset, PNE obtains a large improvement in performance gain with relative 5.0\% HR@5, and 4.2\% NDCG@5, and 3.9\% MRR@5. Since the CFNet component of PNE is a neural CF method (with only one hidden layer), results show the benefit of exploiting unstructured text to alleviate the data sparsity issue faced by CF methods (BPR and MLP).

Second, PNE outperforms the traditional hybrid methods HFT and TBPR on two datasets in terms of three ranking metrics. On Amazon dataset, PNE obtains a significantly large improvement in performance gain with relative 55.0\% HR@5, 28.9\% NDCG@5, and 20.4\% MRR@5. On Cheetah Mobile dataset, PNE still obtains reasonably large improvements with relative 17.5\% HR@10, 1.8\% NDCG@10, and 1.9\% MRR@10. Compared with traditional hybrid methods which integrate the text content using topic modelling or word embeddings, results show the benefit of integrating text information through memory networks (and exploiting the interaction data through neural CF).

\begin{figure*}[]
\centering
\includegraphics[width=14cm, height=7cm]{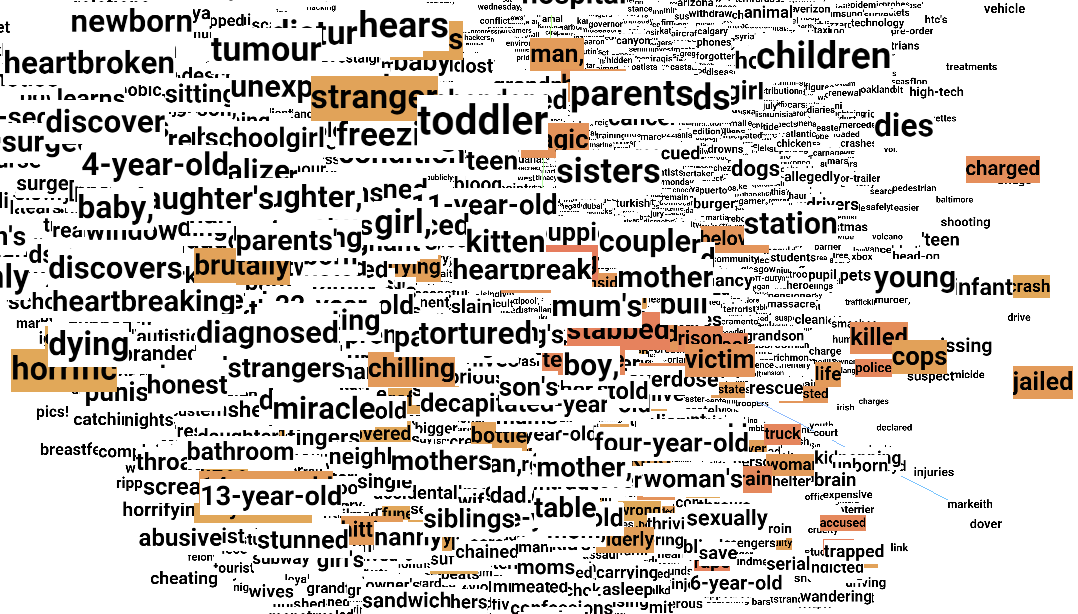}
\caption{Visualization of word embeddings.}
\label{fig:vis_drug}
\end{figure*}

Last, PNE outperforms neural hybrid method LCMR by a large margin on Amazon dataset with relative improvements of 16.2\% HR@5, 9.6\% NDCG, and 7.4\% MRR@5. PNE obtains reasonable improvements on Cheetah Mobile dataset with relative improvements of 3.1\% HR@5, 2.8\% NDCG, and 2.7\% MRR. The design of CFNet of PNE is more reasonable than that of centralized memory module of LCMR which is equivalent to use a softmax activation between two hidden layers. The results show the effectiveness of fusing strategy in PNE to exploit unstructured text via MemNet and the interaction data via CFNet.

\subsection { Analysis }

We first evaluate the effects of the dimensionality of the embedding space. The $x$-axis in Figure~\ref{fig:embedding} is the dimension of user/item and hence the dimensionality of input to CFNet is double since we adopt concatenation. It clearly indicates that the embedding should not be too small due to the possibility of information loss and the limits of expressiveness. The dimension 75 (and hence $d=150$) is a good tradeoff between recommendation performance and computation burden.

We next show optimization curves of performance and loss (averaged over all examples) against iterations on Cheetah Mobile dataset in Figure~\ref{fig:performance-loss-epoch}. The model learns quickly in the first 20 iterations and improves slowly until 50, while training losses continue to go down and valid losses stabilize. The average time per epoch of PNE takes 68.1s and as a reference it is 34.5s for MLP using one NVIDIA TITAN Xp GPU.

\subsection { Visualization }

We visualize learned word embeddings $\bm{A}$. We show that we can learn meaningful semantics for word embeddings such that words are to cluster when they have relevant semantics. We give an example to show the neighbors of the word ``drug'' in the 3D space by projecting the high-dimensional word vectors using TensorFlow~\footnote{\url{https://projector.tensorflow.org/}} as shown in Figure~\ref{fig:vis_drug}. The top nearest neighbors of {\it drug} are: {\it shot, shoots, gang, murder, killing, rape, stabbed, truck, school, police, teenage}. We can see they are highly semantic relevant. We may also infer that school teenagers have close relationships to the drug issue from the Cheetah News corpus. This should raise a concern for the whole society and it shows the society impact of natural language processing~\cite{hovy2016social}. Try it here: {\url{https://www.dropbox.com/sh/ef74fpagf6sd137/AACXF6EnEY6QBdmJcyIW4RE0a?dl=0}}.

\section{Conclusion}

We showed that relational interactions can be effectively integrated with unstructured text under a neural embedding model. Our method attentively focuses relevant words to match user preferences with user and item attentions (semantic factor) and captures nonlinear relationships between users and items (behavior factor). Experiments show better performance than five baselines on two real-world datasets in terms of three ranking metrics. We learn meaningful word embeddings and rethink the society impact of language processing technology.

\noindent
{\bf Acknowledgment} The work is supported by HK CERG projects 16211214/16209715/16244616, NSFC 61673202, and HKPFS PF15-16701.

\bibliography{naaclhlt2019}

\begin{thebibliography}{31}
\expandafter\ifx\csname natexlab\endcsname\relax\def\natexlab#1{#1}\fi

\bibitem[{Abadi et~al.(2016)Abadi, Barham, Chen, Chen, Davis, Dean, Devin,
  Ghemawat, Irving, Isard et~al.}]{abadi2016tensorflow}
Mart{\'\i}n Abadi, Paul Barham, Jianmin Chen, Zhifeng Chen, Andy Davis, Jeffrey
  Dean, Matthieu Devin, Sanjay Ghemawat, Geoffrey Irving, Michael Isard, et~al.
  2016.
\newblock Tensorflow: a system for large-scale machine learning.
\newblock In \emph{OSDI}, volume~16, pages 265--283.

\bibitem[{Bansal et~al.(2016)Bansal, Belanger, and McCallum}]{bansal2016gru}
Trapit Bansal, David Belanger, and Andrew McCallum. 2016.
\newblock Ask the gru: Multi-task learning for deep text recommendations.
\newblock In \emph{Proceedings of the 10th ACM Conference on Recommender
  Systems}, pages 107--114. ACM.

\bibitem[{Bao et~al.(2014)Bao, Fang, and Zhang}]{bao2014topicmf}
Yang Bao, Hui Fang, and Jie Zhang. 2014.
\newblock Topicmf: simultaneously exploiting ratings and reviews for
  recommendation.
\newblock In \emph{Proceedings of the Twenty-Eighth AAAI Conference on
  Artificial Intelligence}, pages 2--8. AAAI Press.

\bibitem[{Catherine and Cohen(2017)}]{catherine2017transnets}
Rose Catherine and William Cohen. 2017.
\newblock Transnets: Learning to transform for recommendation.
\newblock In \emph{Proceedings of the Eleventh ACM Conference on Recommender
  Systems}, pages 288--296. ACM.

\bibitem[{Cheng et~al.(2016)Cheng, Koc, Harmsen, Shaked, Chandra, Aradhye,
  Anderson, Corrado, Chai, Ispir et~al.}]{cheng2016wide}
Heng-Tze Cheng, Levent Koc, Jeremiah Harmsen, Tal Shaked, Tushar Chandra,
  Hrishi Aradhye, Glen Anderson, Greg Corrado, Wei Chai, Mustafa Ispir, et~al.
  2016.
\newblock Wide \& deep learning for recommender systems.
\newblock In \emph{Proceedings of the 1st Workshop on Deep Learning for
  Recommender Systems}, pages 7--10. ACM.

\bibitem[{Dziugaite and Roy(2015)}]{dziugaite2015neural}
Gintare~Karolina Dziugaite and Daniel~M Roy. 2015.
\newblock Neural network matrix factorization.
\newblock \emph{arXiv preprint arXiv:1511.06443}.

\bibitem[{Ebesu et~al.(2018)Ebesu, Shen, and Fang}]{ebesu2018memory}
Travis Ebesu, Bin Shen, and Yi~Fang. 2018.
\newblock Collaborative memory network for recommendation systems.
\newblock In \emph{The 41st International ACM SIGIR Conference on Research and
  Development in Information Retrieval (SIGIR)}, pages 195--204, New York, NY,
  USA. ACM.

\bibitem[{Ganu et~al.(2009)Ganu, Elhadad, and Marian}]{ganu2009review}
Gayatree Ganu, Noemie Elhadad, and Am{\'e}lie Marian. 2009.
\newblock Beyond the stars: improving rating predictions using review text
  content.
\newblock In \emph{International Workshop on the Web and Databases (WebDB)},
  volume~9, pages 1--6.

\bibitem[{He and McAuley(2016)}]{he2016vbpr}
Ruining He and Julian McAuley. 2016.
\newblock Vbpr: visual bayesian personalized ranking from implicit feedback.
\newblock In \emph{Proceedings of the Thirtieth AAAI Conference on Artificial
  Intelligence}, pages 144--150. AAAI Press.

\bibitem[{He et~al.(2017)He, Liao, Zhang, Nie, Hu, and Chua}]{he2017neural}
Xiangnan He, Lizi Liao, Hanwang Zhang, Liqiang Nie, Xia Hu, and Tat-Seng Chua.
  2017.
\newblock Neural collaborative filtering.
\newblock In \emph{Proceedings of the 26th International Conference on World
  Wide Web}, pages 173--182. International World Wide Web Conferences Steering
  Committee.

\bibitem[{Hovy and Spruit(2016)}]{hovy2016social}
Dirk Hovy and Shannon~L Spruit. 2016.
\newblock The social impact of natural language processing.
\newblock In \emph{Proceedings of the 54th Annual Meeting of the Association
  for Computational Linguistics (Volume 2: Short Papers)}, volume~2, pages
  591--598.

\bibitem[{Hu and Dai(2017)}]{hu2017tbpr}
Guang-Neng Hu and Xin-Yu Dai. 2017.
\newblock Integrating reviews into personalized ranking for cold start
  recommendation.
\newblock In \emph{Pacific-Asia Conference on Knowledge Discovery and Data
  Mining}, pages 708--720. PAKDD.

\bibitem[{Hu et~al.(2018{\natexlab{a}})Hu, Dai, Qiu, Xia, Li, Huang, and
  Chen}]{MR3PP}
Guang-Neng Hu, Xin-Yu Dai, Feng-Yu Qiu, Rui Xia, Tao Li, Shu-Jian Huang, and
  Jia-Jun Chen. 2018{\natexlab{a}}.
\newblock Collaborative filtering with topic and social latent factors
  incorporating implicit feedback.
\newblock \emph{ACM Transactions on Knowledge Discovery from Data (TKDD)},
  12(2):23.

\bibitem[{Hu et~al.(2015)Hu, Dai, Song, Huang, and Chen}]{MR3}
Guang-Neng Hu, Xin-Yu Dai, Yunya Song, Shu-Jian Huang, and Jia-Jun Chen. 2015.
\newblock A synthetic approach for recommendation: combining ratings, social
  relations, and reviews.
\newblock In \emph{Twenty-Fourth International Joint Conference on Artificial
  Intelligence}.

\bibitem[{Hu et~al.(2018{\natexlab{b}})Hu, Zhang, and Yang}]{CoNet}
Guangneng Hu, Yu~Zhang, and Qiang Yang. 2018{\natexlab{b}}.
\newblock Conet: Collaborative cross networks for cross-domain recommendation.
\newblock In \emph{Proceedings of the 27th ACM International Conference on
  Information and Knowledge Management}, pages 667--676. ACM.

\bibitem[{Hu et~al.(2018{\natexlab{c}})Hu, Zhang, and Yang}]{hu2018lcmr}
Guangneng Hu, Yu~Zhang, and Qiang Yang. 2018{\natexlab{c}}.
\newblock Lcmr: Local and centralized memories for collaborative filtering with
  unstructured text.
\newblock \emph{arXiv preprint arXiv:1804.06201}.

\bibitem[{Huang et~al.(2017)Huang, Zhang, and Huang}]{huang2017mention}
Haoran Huang, Qi~Zhang, and Xuanjing Huang. 2017.
\newblock Mention recommendation for twitter with end-to-end memory network.
\newblock In \emph{Proceedings of the 26th International Joint Conference on
  Artificial Intelligence}, pages 1872--1878. AAAI Press.

\bibitem[{Kingma and Ba(2015)}]{Kingma2015AdamAM}
Diederik~P. Kingma and Jimmy Ba. 2015.
\newblock Adam: A method for stochastic optimization.
\newblock \emph{International Conference on Learning Representations}.

\bibitem[{Koren(2008)}]{koren2008mf}
Yehuda Koren. 2008.
\newblock Factorization meets the neighborhood: a multifaceted collaborative
  filtering model.
\newblock In \emph{Proceedings of the 14th ACM SIGKDD international conference
  on Knowledge discovery and data mining}, pages 426--434. ACM.

\bibitem[{Koren et~al.(2009)Koren, Bell, and Volinsky}]{koren2009matrix}
Yehuda Koren, Robert Bell, and Chris Volinsky. 2009.
\newblock Matrix factorization techniques for recommender systems.
\newblock \emph{IEEE Computer}, (8):30--37.

\bibitem[{Liu et~al.(2018)Liu, Wei, Zhang, Yan, and Yang}]{liu2018transferable}
Bo~Liu, Ying Wei, Yu~Zhang, Zhixian Yan, and Qiang Yang. 2018.
\newblock Transferable contextual bandit for cross-domain recommendation.
\newblock In \emph{AAAI Conference on Artificial Intelligence,}.

\bibitem[{McAuley and Leskovec(2013)}]{mcauley2011hft}
Julian McAuley and Jure Leskovec. 2013.
\newblock Hidden factors and hidden topics: understanding rating dimensions
  with review text.
\newblock In \emph{Proceedings of the 7th ACM conference on Recommender
  systems}, pages 165--172. ACM.

\bibitem[{Mnih and Salakhutdinov(2008)}]{mnih2008pmf}
Andriy Mnih and Ruslan~R Salakhutdinov. 2008.
\newblock Probabilistic matrix factorization.
\newblock In \emph{Advances in neural information processing systems}, pages
  1257--1264.

\bibitem[{Pennington et~al.(2014)Pennington, Socher, and
  Manning}]{Pennington2014GloveGV}
Jeffrey Pennington, Richard Socher, and Christopher Manning. 2014.
\newblock Glove: Global vectors for word representation.
\newblock In \emph{Proceedings of the 2014 conference on empirical methods in
  natural language processing (EMNLP)}, pages 1532--1543.

\bibitem[{Rendle et~al.(2009)Rendle, Freudenthaler, Gantner, and
  Schmidt-Thieme}]{rendle2009bpr}
Steffen Rendle, Christoph Freudenthaler, Zeno Gantner, and Lars Schmidt-Thieme.
  2009.
\newblock Bpr: Bayesian personalized ranking from implicit feedback.
\newblock In \emph{Proceedings of the twenty-fifth conference on uncertainty in
  artificial intelligence}, pages 452--461. AUAI Press.

\bibitem[{Sarwar et~al.(2001)Sarwar, Karypis, Konstan, and
  Riedl}]{sarwar2001item}
Badrul Sarwar, George Karypis, Joseph Konstan, and John Riedl. 2001.
\newblock Item-based collaborative filtering recommendation algorithms.
\newblock In \emph{Proceedings of the 10th international conference on World
  Wide Web}, pages 285--295. ACM.

\bibitem[{Tay et~al.(2018)Tay, Anh~Tuan, and Hui}]{tay2018latent}
Yi~Tay, Luu Anh~Tuan, and Siu~Cheung Hui. 2018.
\newblock Latent relational metric learning via memory-based attention for
  collaborative ranking.
\newblock In \emph{Proceedings of the 2018 World Wide Web Conference on World
  Wide Web}, pages 729--739. International World Wide Web Conferences Steering
  Committee.

\bibitem[{Wang and Blei(2011)}]{wang2011ctr}
Chong Wang and David~M Blei. 2011.
\newblock Collaborative topic modeling for recommending scientific articles.
\newblock In \emph{Proceedings of the 17th ACM SIGKDD international conference
  on Knowledge discovery and data mining}, pages 448--456. ACM.

\bibitem[{Wang et~al.(2015)Wang, Wang, and Yeung}]{wang2015cdl}
Hao Wang, Naiyan Wang, and Dit-Yan Yeung. 2015.
\newblock Collaborative deep learning for recommender systems.
\newblock In \emph{Proceedings of the 21th ACM SIGKDD International Conference
  on Knowledge Discovery and Data Mining}, pages 1235--1244. ACM.

\bibitem[{Zhang et~al.(2016)Zhang, Yuan, Lian, Xie, and Ma}]{zhang2016cke}
Fuzheng Zhang, Nicholas~Jing Yuan, Defu Lian, Xing Xie, and Wei-Ying Ma. 2016.
\newblock Collaborative knowledge base embedding for recommender systems.
\newblock In \emph{Proceedings of the 22nd ACM SIGKDD international conference
  on knowledge discovery and data mining}, pages 353--362. ACM.

\bibitem[{Zheng et~al.(2017)Zheng, Noroozi, and Yu}]{zheng2017joint}
Lei Zheng, Vahid Noroozi, and Philip~S Yu. 2017.
\newblock Joint deep modeling of users and items using reviews for
  recommendation.
\newblock In \emph{Proceedings of the Tenth ACM International Conference on Web
  Search and Data Mining}, pages 425--434. ACM.

\end{thebibliography}
\bibliographystyle{acl_natbib}

\end{document}